\documentclass[floatfix,preprint,aip,rsi]{revtex4-2} 
\usepackage{lipsum}
\usepackage{lineno}
\usepackage{graphicx}
\usepackage{amsmath}
\usepackage{epsfig}
\usepackage{subeqnarray}
\usepackage{color}
\usepackage{bbold}
\usepackage{dsfont}
\usepackage{tikz}
\usepackage{cancel}
\usetikzlibrary{positioning,calc}
\usepackage[version=3]{mhchem}
\usepackage{color}
\usepackage[normalem]{ulem}

\newcommand{\mt}[1]{\mathrm{#1}}

\newcommand{\fig}[1]{Figure~\ref{#1}}
\newcommand{\eq}[1]{Eq.~\ref{#1}}

\newcommand{\ie}{\textit{i.e.},~}
\newcommand{\via}{\textit{via~}}

\newcommand{\cm}{$\mt{cm^{-1}}$}
\graphicspath{{./Figures/}}

\begin{document}

\title{Sparse Infrared Spectroscopy for Detection of Volatile Organic Compounds}

\author{Mira Welner}
\author{Andre Hazbun}
\author{Thomas E. Beechem}
\email{tbeechem@purdue.edu}
\affiliation{School of Mechanical Engineering and Birck Nanotechnology Center, Purdue University, West Lafayette, 47907, IN, USA}
\date{\today}

\begin{abstract}
To reduce the complexity of infrared spectroscopy hardware while maintaining performance, a data informed, task-specific, spectral collection approach termed Sparse Infrared Spectroscopy (SIRS) is developed.  Using a numerically based virtual experiment based on a quantitatively accurate infrared database, non-negative matrix factorization is used to identify the spectral pass bands of a minimal number of filters necessary to identify volatile organic compounds (VOC) within either an inert background or mixture of gases. The data-driven approach is found capable of identifying contaminants at the 1-10 part per million level (PPM) with $\mathrm{\sim~20-50}$ spectral samples as opposed to the more than 1,000 typical of a traditional infrared spectrum. Reasonably robust to both noise and the characteristics of the base compound in a mixture, the task-specific spectral sampling points to simplified hardware design that maintains performance. 
\end{abstract}
\maketitle

\section{Introduction}
Infrared (IR) spectroscopy is commonly employed for effectively and non-destructively detecting harmful components in solids, liquids, and gases. By analyzing to what extent different infrared wavelengths are absorbed, reflected or emitted by a substance, materials are identified through their so-called molecular ``fingerprint."  The ability to fingerprint provides the technique's efficacy in detecting toxins in food \cite{freitag_2022} and water,\cite{zulkifli_2018} as well as identifying pollutants in air.\cite{malachowski_1994} Industrially, IR tools are frequently used to monitor semiconductor processes,\cite{rosenthal_1998,bec_2021} the by-products of those processes \cite{sung_2014a,sung_2014,tsao_2015} and even as a tool to enable sustainable semiconductor manufacturing.\cite{cherniienko_2024}  Here, we pursue methods to leverage these advantages while minimizing hardware requirements through the development of a data informed spectral collection approach termed Sparse Infrared Spectroscopy (SIRS).

Despite the efficacy of IR-spectroscopy, the size, weight, power and cost (SWAP-C) of its equipment severely restricts the applications for which it is employed. The use of large diffractive optics or interferometers have intrinsically tied the performance of these tools to their physical size while also requiring either a multi-pixel imaging array such as that used in NIR spectrometers or frequent movement inherent to the use of an interferometer. Even still, progress has been made in miniaturizing IR-spectroscopic equipment. For instance, microfabricated Michelson,\cite{mortada_2022} Fabry–Pérot,\cite{correia_1999} and Mach-Zender\cite{souza_2018} interferometers have each been demonstrated having significant reductions in footprint relative to standard tools.

While hardware changes reduce the physical dimensions of these systems, they tend to increase both cost and complexity.  This, in turn, limits the applications for which can be employed. Making IR-spectroscopy more accessible requires wholesale changes in the system's very architecture rather than just its miniaturization.  For this reason, filter-based spectroscopic approaches have been increasingly pursued.\cite{yang_2021} In this approach, neither the diffraction nor interference of light is used to differentiate between wavelengths.  Rather, discrete filters having purposely designed ``pass-bands" are placed in front of a detector. \textit{A priori} knowledge of these pass bands allows the spectrum to be directly,\cite{correia_1999,tittl_2018} or computationally,\cite{cerjan_2019a,oliver_2012} reconstructed. Performance of these systems is often tied to the number of filters or the number of sensing elements, which again scales complexity with performance for general tasks.

Beyond the laboratory bench, spectrometers do not need to perform in every wavelength range, or detect every variety of substance. Rather, their use is tied to specific tasks. This might be, for example, identifying a particular chemical, pollutant, or industrial by-product.\cite{ni_2023a}  Alternatively, the tool might be used for classification to differentiate between a set of known compounds.\cite{ng_2009a} Performing these tasks does not require collection of a full spectrum. Instead, only those signals related to the task must be detected. Thus, with a defined task in mind, performance becomes decoupled from complexity.  This presumes, however, that a minimum number of useful spectral samples necessary to perform the spectral task can be identified. 

To this end, we employ a data-driven methodology employing non-negative matrix factorization (NMF) to identify a minimal number of spectral samples necessary to perform a narrow, but well-defined, spectroscopy task.  The task itself centers on identifying a volatile organic compound (VOC) within either an inert background or mixture of gases. Using a numerically based ``virtual experiment"  based on a quantitatively accurate infrared database,\cite{chu_1999} the data-driven approach is found capable of identifying contaminants at the 1-10 part per million level (PPM) with $\mathrm{\sim~20-50}$ spectral samples as opposed to the more than 1,000 typical of most IR-spectra. This data-driven, task specific, approach provides a design path to reduce the complexity of IR-hardware and thus facilitate its more widespread use. 

\section{Method and Approach}
\begin{figure}[tb]
\centering\includegraphics[width=\linewidth]{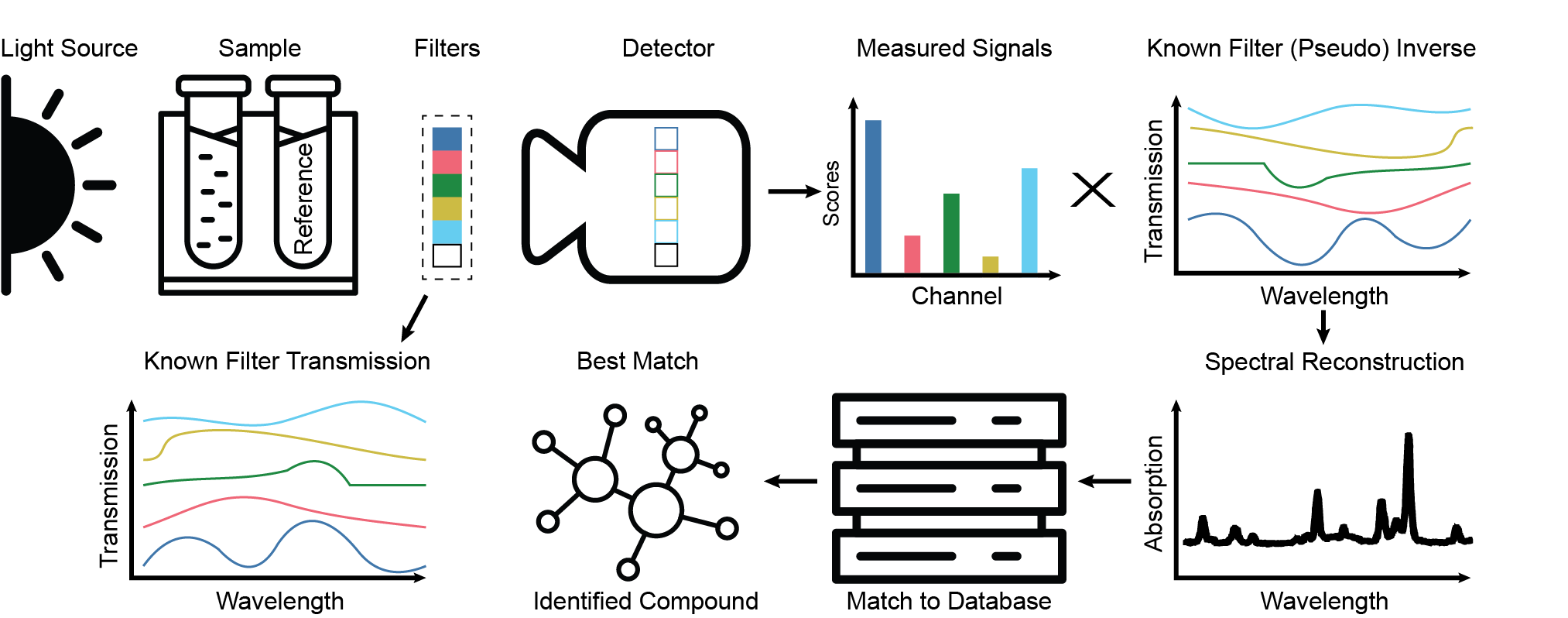}
\caption{Conceptual representation of task-specific sparse infrared spectroscopy (SIRS).  A broadband IR-source illuminates a sample and reference after which the light is directed through filters having pass bands matching the basis spectra of the non-negative matrix factorization of the underlying dataset.  The resulting signals represent scores, or weights, that when multiplied by the pseudo-inverse of the underlying basis spectra reconstruct the spectra.  By comparing the reconstructed spectra to the underlying database, compounds are identified by measuring 10’s of channels rather than 1000's.}
\label{fig_1_schematic}
\end{figure}

A virtual experiment is built presuming a filter-based spectroscopy system like that shown schematically in \fig{fig_1_schematic} designed to differentiate between the 40 separate VOC compounds within the NIST Quantitative IR-database.\cite{chu_1999} Like typical NDIR hardware, a broadband infrared light source illuminates a sample and measures absorption relative to a known reference.  Spectral information is gathered by passing the transmitted light through a few-element (\ie $20\lesssim n \lesssim 50$) filter array that is matched to a complementary detector having the same number of channels.  Importantly, transmittance through each filter is general over the spectral range of interest and chosen to perform a particular spectral task. This is in contrast to the uniform pass band used in standard spectrometers.  The measured signal is a set of $n$ scores corresponding to the amount of light passing through each filter. The spectral task seeks to minimize the number of $n$ scores needed to discriminate between the 40 separate VOCs included in the database as a function of their concentration and mixing within a background $N_2$-gas.\cite{chu_1999} The following will explain how this is task is accomplished.

\begin{figure}[tb]
\centering\includegraphics[width=0.8\linewidth]{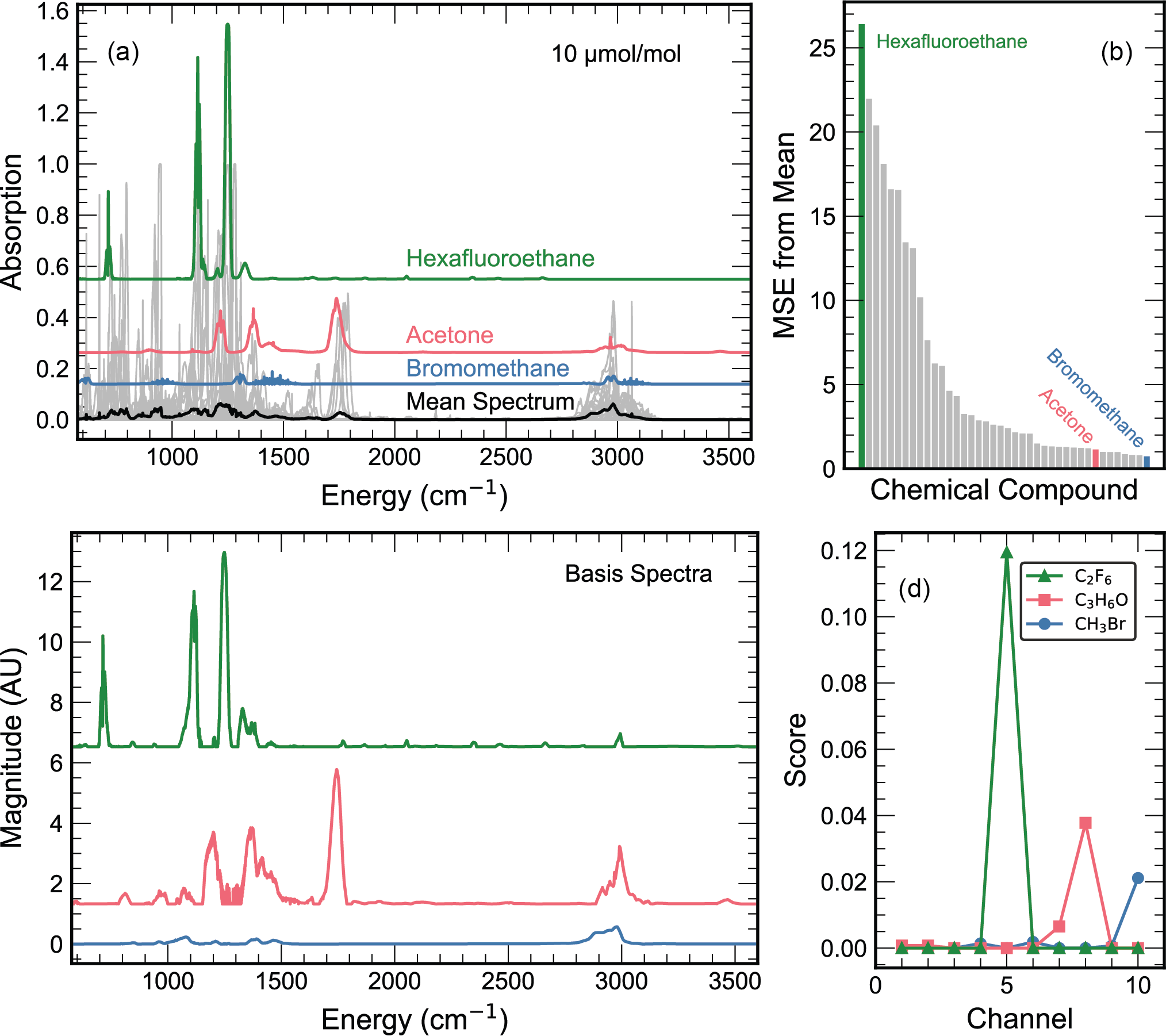}
\caption{(a) Simulated absorption for the 40 separate volatile organic compounds (VOC) when passing through a 20 m path length at a concentration of 10 µmol/mol, as shown by the thin gray lines.  The mean spectrum of the database is accompanied by bromomethane, acetone, and hexafluoroethane used as case studies.  These spectra are offset for clarity.  (b) Difference in spectral response relative to mean of database for each of the compounds. (c) Basis spectrum of channel scoring highest for the highlighted compounds. (d) Score spectra for each of the compounds when utilizing 10 channels. }
\label{fig_2_spectra}
\end{figure}

The NIST quantiative infrared database tabulates infrared absorption coefficients ($\alpha_i$) for the included VOCs across the molecular fingerprinting energy range spanning 576 \cm~to 3974 \cm. These absorption coefficients allow calculation of absorption ($A$) presuming a given concentration ($c_i$) of the VOC within a background of \ce{N_2} gas after light has traveled a distance ($L$) through the mixture using Beer's Law:\cite{hecht_2016}
\begin{equation}
\label{eq_Beer}
\log_{10}(A)=-L\sum (\alpha_i \cdot c_i)
\end{equation}
where the summation takes place across all VOCs present in the mixture.   \fig{fig_2_spectra}(a) displays the resulting IR-absorption spectrum from each of the compounds in the database assuming a concentration of 10 µmol of VOC per mole of \ce{N_2} carrier gas (\ie $c_i$= 10 µmol/mol).  The mean of the database as a whole overlays these spectra. In addition, the IR-absorption spectrum of hexafluoroethane (\ce{C_2F_6}), acetone (\ce{C_3H_6O}), and bromomethane (\ce{CH_3Br}) are highlighted because of the large difference in their respective variances from the mean spectrum, as highlighted in \fig{fig_2_spectra}(b).  Due to these differences in variance from the mean spectrum, the three compounds were used as case studies to assess the general capability of the approach.

Qualitative examination of the database highlights the inherent sparseness of spectral data.  Three broad bands\textemdash 500-1500 \cm, 1500-2000 \cm, and 2700-3200 \cm\textemdash contain the vast majority of absorption.  Sampling from regions outside of these bands contains little information and is, therefore, unnecessary.  This sparseness, in turn, motivates data-driven methods to identify the  minimal number of information-rich measurements to perform the needed spectral task (\ie identifying VOCs) so that the hardware need can be minimized as well.  

Non-negative matrix factorization (NMF) is used for this purpose. NMF is a linear dimensionality reduction method that capitalizes upon the constraint that spectral data is inherently non-negative to approximate a dataset in a manner that enhances sparsity.\cite{gillis_2014} In doing so, NMF also highlights how to design a minimal number of information-rich filters. 

To verify this assertion, a synthetic database was constructed by calculating the IR-absorption spectrum for each of the 40 VOCs in the original NIST dataset at 16 different concentrations ranging from 0.1 to 100 µmol/mol presuming a path length of 20 m, which is typical of commercial long-path gas cells. The resulting database ($\mathbf{V}$) is made up of $R$ rows corresponding to each of the 3526 spectral energies sampled in the original dataset and $C$ columns tabulating the absorption at each of these energies for the $40 \times 16=625$ separate simulated spectra. Utilizing NMF, $\mathbf{V}$ is approximated by two non-negative matrices, termed here the scores ($\mathbf{S}$), and basis spectra ($\mathbf{H}$), such that $\mathbf{V}\approx\mathbf{SH}$. $\mathbf{S}$ and $\mathbf{H}$ have sizes of $R\times \delta$ and $\delta \times C$, respectively, where $\delta$ represents an arbitrary number of dimensions or, to emphasize the link with the physical measurement, data channels. From a more mathematical perspective, the basis spectra ($\mathbf{H}$) represent the non-negative principal components of the dataset while the scores ($\mathbf{S}$) indicate ``how much" of each basis is needed to approximate any spectrum in the database. Practically, the implementation of NMF used the Scikit-learn Python library using an alternating minimization method working in concert with the Frobenius norm using a coordinate descent solver and random initialization of the fitting.\cite{pedregosa_2011} These algorithmic choices are not necessarily optimal but proved sufficiently capable for demonstration of the approach. 

\section{Results and Discussion}
\fig{fig_2_spectra}(c) and (d) highlight NMF's ability to identify spectrally relevant features while maintaining sparseness. For example, the ``score spectra" of \fig{fig_2_spectra}(d) obtained for hexafluoroethane, acetone, and bromoethane in a 10-channel approximation each have a strong maximum accompanied by several near zero readings.  The corresponding basis spectra of these high-scoring channels shown in \fig{fig_2_spectra}(c), meanwhile, exhibit clear similarities with their corresponding parent spectra [see \fig{fig_2_spectra}(a)] even as they do not map in a 1 to 1 manner.  

\begin{figure}[tb]
\centering\includegraphics[width=\linewidth]{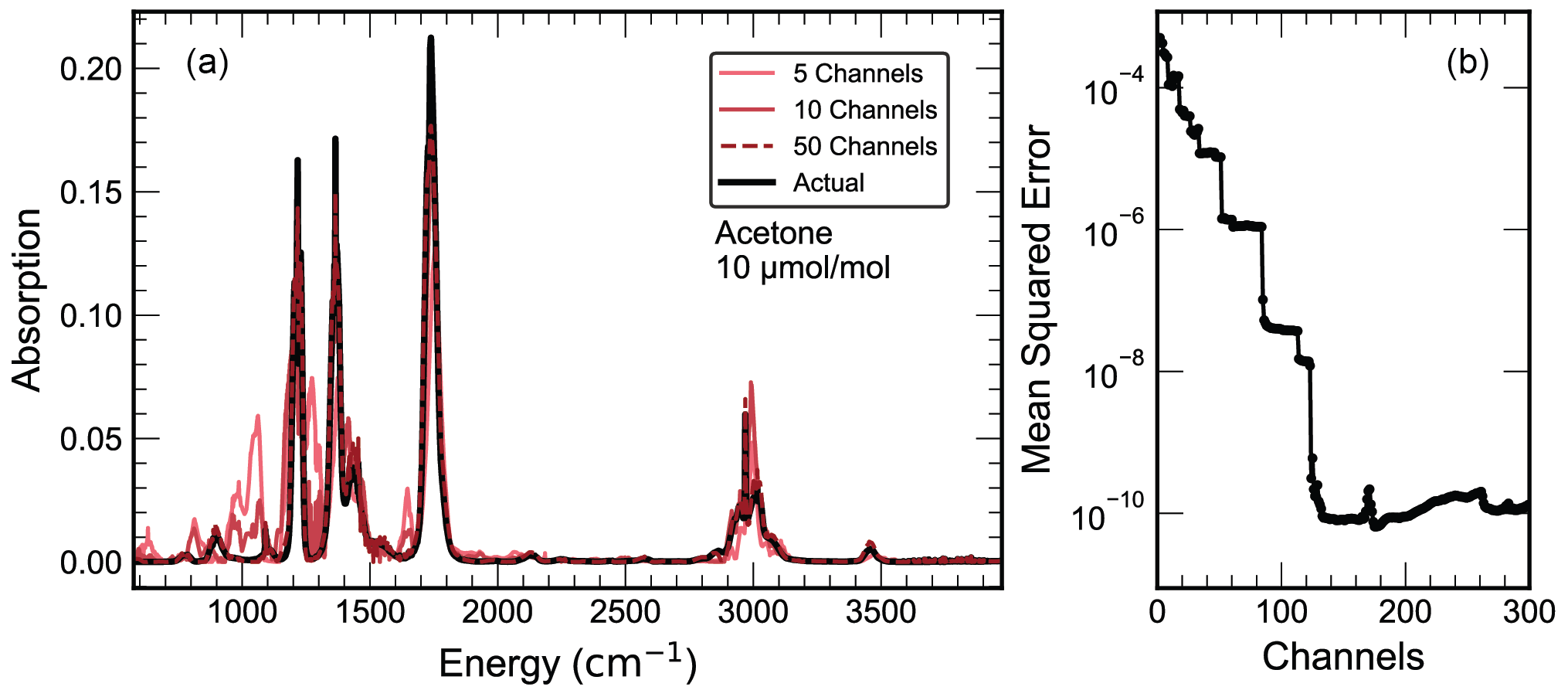}
\caption{(a) Evolution of reconstructed spectrum of acetone with channel number. (b) The evolution of MSE quantitatively shows how the reconstruction converges rapidly to the actual response with just 5-50 channels.  This suggests that only a few pixels may be needed to discriminate between compounds.}
\label{fig_3_acetone}
\end{figure} 
The sparseness created by the NMF, in turn, means that only a few channels are needed to make reasonable approximations of the underlying spectra. This is exhibited in \fig{fig_3_acetone}(a) where the strongest IR-absorption features of acetone are accurately approximated with just 5 channels while with 50 channels the actual spectrum is almost perfectly recreated. Additional channels provide only comparatively small improvements as seen by the evolution in mean squared error (MSE) in \fig{fig_3_acetone}(b). Altogether, these results show that NMF can re-cast a complicated spectral database in a sparse and compressed manner thereby opening a path to performing spectral identification with just a few measurements. 

To understand how NMF enables completion of this spectral task, consider light incident on the filters in \fig{fig_1_schematic} after moving through the sample having an unknown absorption, $A(\lambda)$. Each filter has a separate, and disparate, spectral transmittance that matches one of the spectral bases identified by NMF like that shown in \fig{fig_2_spectra}(c).  Thus,  each row of $\mathbf{H}$ can be represented by a single filter having a transmittance given by $H_i(\lambda)$.  Presuming uniform broadband detection of all wavelengths, the detector element matching this filter will read out a value proportional to the total intensity of light irradiating it over all wavelengths. This total can be represented as:
\begin{equation}
\label{eq_Score}
S_i = \int H_{i}(\lambda)A(\lambda) d\lambda \approx \sum_{\lambda_j} H_{i,j}(\lambda_j)A(\lambda_j)
\end{equation}
The integral of \eq{eq_Score} signifies the actual light impinging on a detector pixel after passing through a given filter. Summation is a more useful representation, however, because spectral reconstruction necessitates prior knowledge of the filter transmittance. Filter transmittance, in turn, can be known only at some level of discreteness.  Practically, this will be determined by those wavelengths sampled during characterization of the filters prior to their insertion into the set-up of \fig{fig_1_schematic}.

The discrete nature of any practical measurement enables the resulting score spectra, $\mathbf{S}$, to be described in matrix form through
\begin{equation}
\label{eq_Smatrix}
\mathbf{S} = \mathbf{HA}
\end{equation}
where the unknown, wavelength-dependent, absorption is now represented by a column vector.  Implicit in this definition is that $\mathbf{A}$ has a size equal to the number of wavelengths sampled during filter characterization, which is typically of $\mathcal{O}(10^3)$. Since the transmittance of all filters is known, the pseudo-inverse of the total filter set, $\mathbf{H}^+$, can be calculated. Spectral reconstruction is thus accomplished via 
\begin{equation}
\label{eq_reconstruction}
\mathbf{A_R = H^+S}.
\end{equation}
where $\mathbf{A_R}$ is the reconstructed spectrum that approximates $\mathbf{A}$. Note that while the reconstruction will have $\mathcal{O}(10^3)$ values, only $\delta$ measurements [$\mathcal{O}(10^1)$] must be collected.  Furthermore, the reconstruction itself must not be perfect but only sufficient to correctly perform the spectral task of differentiating between the 40 VOCs of the NIST database. The goal is to minimize the number of channels necessary to successfully perform this spectral task.

To this end, we quantified the number of channels needed to correctly identify each of the VOCs as a function of their concentration within a medium of nitrogen.  The virtual experiment was simulated by first calculating the absorption spectra for each of the VOCs at a given concentration via \eq{eq_Beer}. Noise ($\Delta$) was simulated by randomly sampling from a standard Gaussian distribution and modifying the calculated absorption at a given wavelength ($A_{\lambda}$) with the sampled noise value ($\Delta_{\lambda}$) at a given wavelength using the relation
\begin{equation}
\label{eq_noise}
A'_{\lambda} = |A_{\lambda}(1+P\Delta_{\lambda})|
\end{equation}
where $A'_{\lambda}$ is the noisy signal and $P$ is a proportional value set to range from $0-0.01$ for a given simulation.  An upper limit of $1\%$ noise was chosen as it is easily attainable in standard FTIR measurements.\cite{chu_1999} 

The noisy absorption was then used to simulate an effective score spectra using \eq{eq_Smatrix}, where $\mathbf{H}$ was determined via the NMF process outlined above. \eq{eq_reconstruction} then reconstructed the spectrum, which was compared to the synthetic database. The contaminant giving rise to the spectrum in the synthetic database having the minimum mean square difference relative to that reconstructed was then taken as the identified compound.  This data flow is shown schematically in \fig{fig_1_schematic}. The minimum number of channels necessary to achieve a certain accuracy in identifying each of the compounds in the database was then quantified.

\begin{figure}[tb]
\centering\includegraphics[width=88 mm]{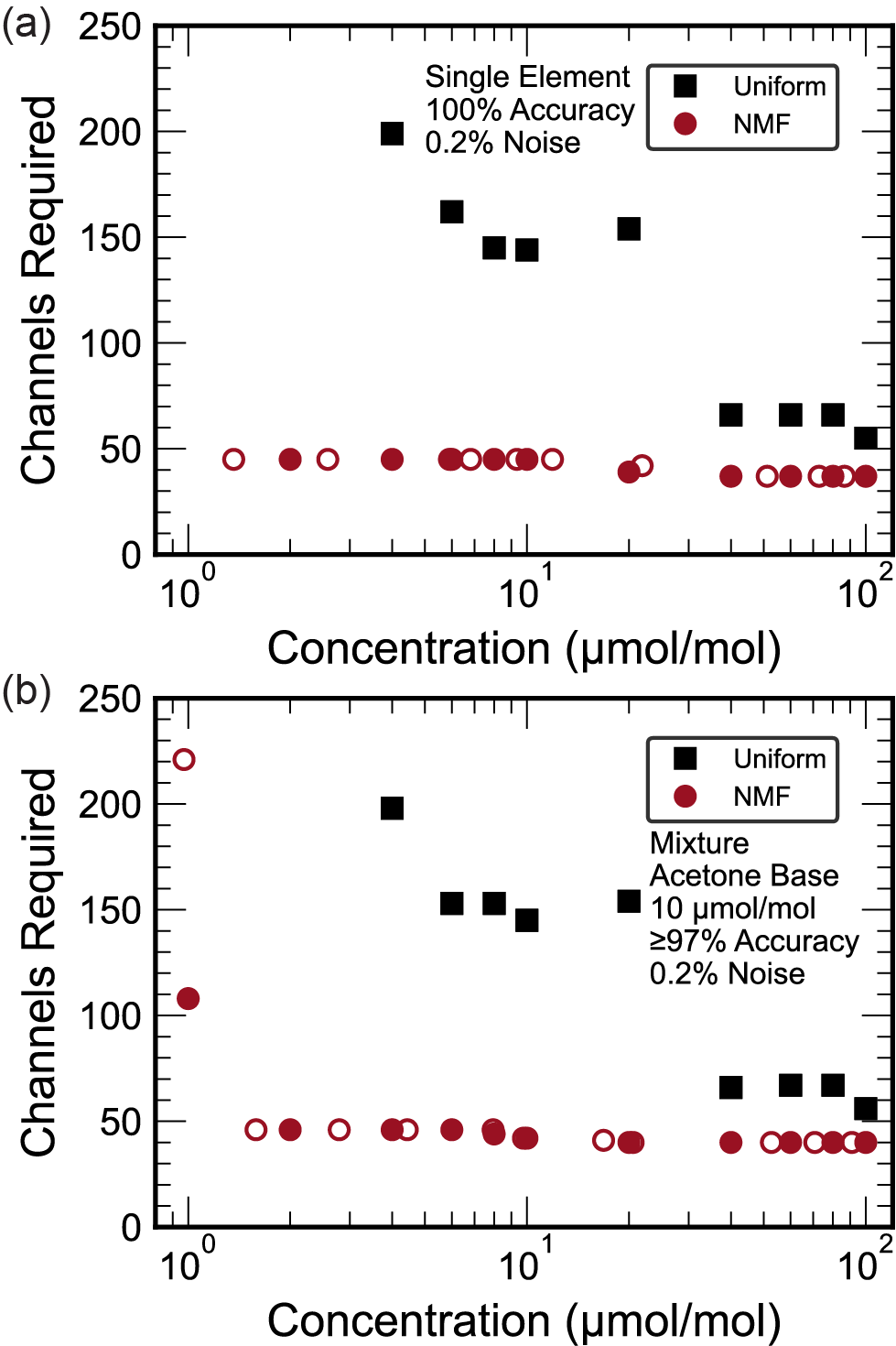}
\caption{(a) Minimum number of channels necessary to correctly identify each VOC within a nitrogen medium as a function of their concentration at a noise level of 0.2$\%$. (b) Required channels needed to identify a secondary component in a known acetone stream of 10 $\mathrm{\mu mol/mol}$ with $97\%$ accuracy.}
\label{fig_4_mindim}
\end{figure} 

As seen in \fig{fig_4_mindim}, the sparse filtering approach is capable of distinguishing between the VOC's with $100 \%$ accuracy using $<50$ channels at concentrations down to $< 2 \; \mathrm{\mu mol/mol}$ even in the presence of non-negligible noise of $0.02\%$. This is, in contrast, to a standard approach utilizing uniformly spaced Gaussian pass bands where nearly $3x$ as many data channels are needed to achieve the same accuracy at $10 \; \mathrm{\mu mol/mol}$. Additionally, the lower-limit of detection is nearly $4x$ greater using the standard, uniform, sampling compared to the sparse approach.  It is also noteworthy that the SIRS technique is capable of interpolation between concentrations included in the NMF-based design of the filters.  Open markers in \fig{fig_4_mindim}, for example, represent concentrations where filter design was blind.  Performance is nearly identical for these ``blind" concentration values as those ``seen" by NMF when designing the filters.  

SIRS is also capable of identifying a contaminant in a known VOC stream.  \fig{fig_4_mindim}(b) plots the minimum number of channels needed to accurately identify $\geq 97\%$ of contaminants (\ie $\leq$1 incorrect identification) within a mixture of a base carrying compound\textemdash in this case acetone at a level of $10 \; \mathrm{\mu mol/mol}$\textemdash as a function of the secondary component's concentration.  Again, the sparse sampling approach outperforms standard spectral sampling of typical spectrometers.  Whereas the standard approach is incapable of identifying the contaminant at concentrations $\lesssim 80\%$ of the base acetone level even when using hundreds of channels, the SIRS approach is much more sensitive.  Reasonable accuracy is apparent at levels down to nearly $10\%$ of the base acetone using $\leq 50$ channels.  Taken together, these results point to the potential of leveraging a sparse spectral sampling approach for a classification task within both single-component streams as well as mixtures. 

\begin{figure}[tb]
\centering\includegraphics[width=88 mm]{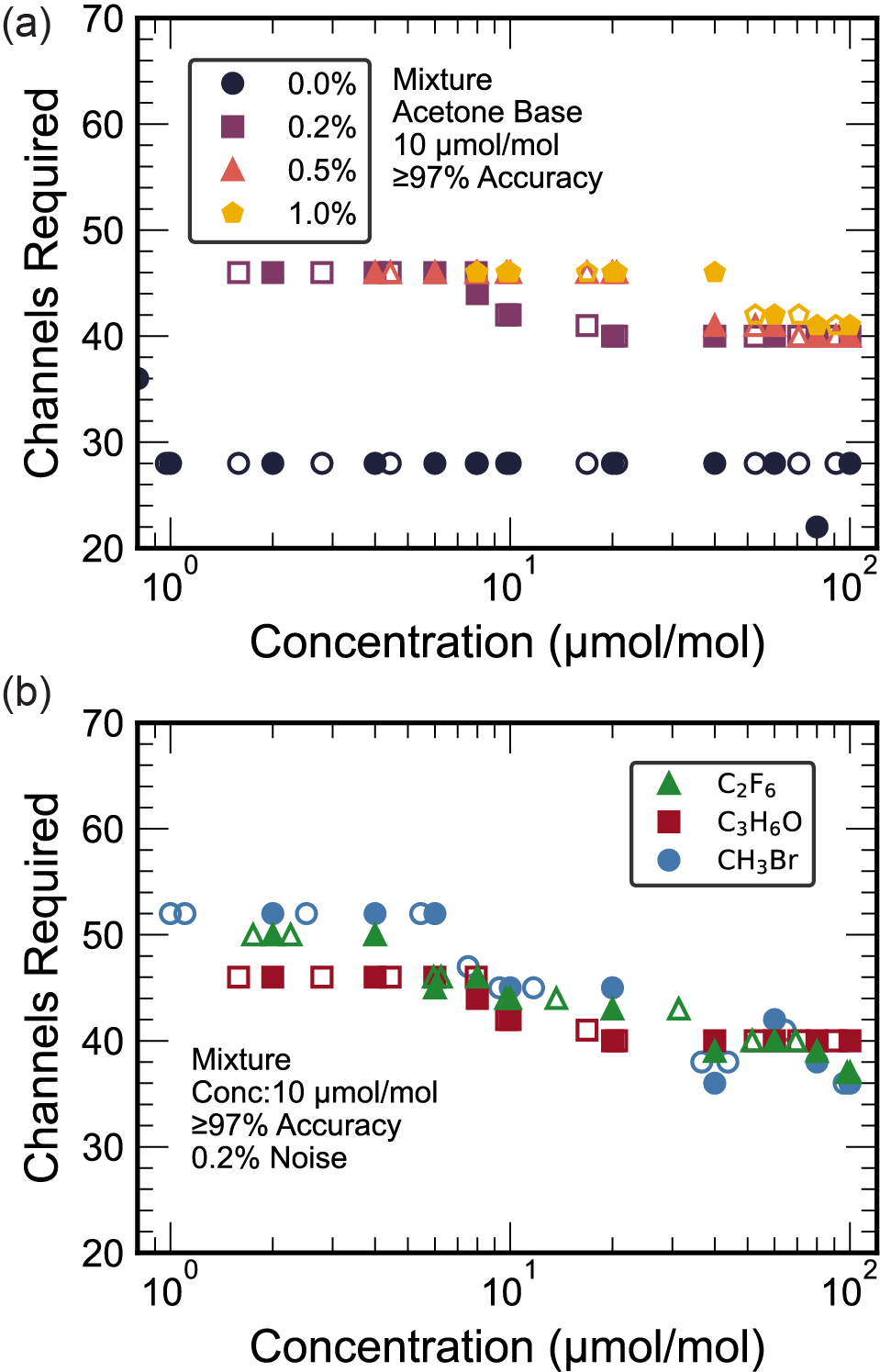}
\caption{(a) Required channels needed to identify a secondary component in a known acetone stream of 10 $\mathrm{\mu mol/mol}$ with $97\%$ accuracy at increasing levels of noise. (b) Number of channels needed to identify a secondary component in a known stream of either hexafluoroethane (\ce{C_2F_6}), acetone (\ce{C_3H_6O}), and bromomethane (\ce{CH_3Br}) at a noise level of 0.2$\%$.}
\label{fig_5_noisebase}
\end{figure} 

SIRS maintains reasonable robustness to noise and the characteristics of the base compound when analyzing mixtures.  Noise, for example, does not acutely affect the number of channels needed to identify a contaminant within a base acetone mixture, as can be seen in \fig{fig_5_noisebase}(a).  Rather, the detection limit is most heavily impacted with the high-noise ($1\%$) condition having a detection limit of 8 µmol/mol relative to the $<$2 µmol/mol level obtainable with a low-noise ($0.2\%$) measurement.  

Consistent performance is also observed when the base compound changes.  \fig{fig_5_noisebase}(b) plots the channel requirement when the base compound is varied between hexafluoroethane (\ce{C_2F_6}), acetone (\ce{C_3H_6O}), and bromomethane (\ce{CH_3Br}).  These compounds provide differing degrees of variance relative to the mean spectrum of the database with hexafluoroethane exhibiting the greatest difference and bromomethane the least [see \fig{fig_2_spectra}(b)]. Irrespective of the base compound's characteristics relative to the database as a whole, similar amounts of data must be collected with 40-50 channels required to meet 97$\%$ accuracy at $0.2\%$ noise.  The lower-limit of detection depends only slightly with variance of the base compound as levels of $\leq$ 2 µmol/mol could be sensed in each case. Ultimate sensitivity is correlated to the base compound.  Bromomethane\textemdash the compound most similar to the mean spectrum of the total database\textemdash exhibits a lower-limit of $\sim$1 µmol/mol while hexafluoroethane\textemdash least similar to mean spectrum\textemdash reaches only slightly less than 2 µmol/mol. Taken in aggregate, the relative consistency in performance with differing base compounds and noise levels suggest the generality of the SIRS approach. 

\section{Conclusions}
Infrared spectroscopy remains both hardware and data intensive. Standard implementations require dispersive or interferometric optical components that collect thousands of data points per spectrum.  Most spectral tasks do not require this level of hardware or data.  Rather, filter-based spectroscopy opens a path to leaner methods \via the use of task-specific spectral sampling. To this end, non-negative matrix factorization was used to identify a minimal number of filters needed to differentiate between a series of volatile organize compounds (VOCs) as a function of their concentration within either an inert background or mixture of gases. Requirements for data collection dropped by an order of magnitude necessitating collection of only 20-50 channels compared to $\gtrsim$1000 typical of traditional approaches even as a $\leq$ 2 µmol/mol lower-limit of detection was attained.  Comparably robust to noise and the characteristics of the base compounds when analyzing mixtures, this sparse infrared spectroscopic (SIRS) approach opens a path to simplified hardware design that maintains performance for an identified task. 


\section{Acknowledgements}
This work was supported in part by Sandia National Laboratories.  Sandia National Laboratories is a multimission laboratory managed and operated by National Technology $\&$ Engineering Solutions of Sandia, LLC, a wholly owned subsidiary of Honeywell International Inc., for the U.S. Department of Energy’s National Nuclear Security Administration under contract DE-NA0003525.

\section{Data Availability}
The data that support the findings of this study are available from the corresponding author upon reasonable request.

\bibliography{Library_Updated}

\end{document}